\documentclass[12pt]{article}

\usepackage{amsmath}
\usepackage{cite}
\usepackage{graphicx}

\title{Inelastic effects\\ in molecular junction transport:\\ 
       Scattering and self-consistent calculations\\ 
       for the Seebeck coefficient}

\author{{\bfseries Michael~Galperin${}^{1}$,
                   Abraham Nitzan${}^{2}$,
                   and Mark~A.~Ratner${}^{1}$}\\*[0.3cm]
 ${}^1$ {\it Department of Chemistry and Materials Research Center,}\\
        {\it Northwestern University, Evanston, IL 60208}\\*[0.2cm]
 ${}^2$ {\it School of Chemistry, The Sackler Faculty of Science,}\\
        {\it Tel Aviv University, Tel Aviv 69978, Israel}
}

\date{\today}

\begin{document}

\maketitle
\thispagestyle{empty}

\newpage
\begin{abstract}
The influence of molecular vibration on the Seebeck coefficient is studied 
within a simple model. Results of a scattering theory approach are compared to 
those of a full self-consistent nonequilibrium Green's function scheme.  
We show, for a reasonable choice of parameters, 
that inelastic effects have non-negligible influence on the
resulting Seebeck coefficient for the junction. 
We note that the scattering theory approach 
may fail both quantitatively and qualitatively. Results of calculation with 
reasonable parameters are in good agreement with recent 
measurements [P.~Reddy~et~al., Science \textbf{315}, 1568 (2007)]
\end{abstract}

\section{\label{intro}Introduction}
Development of experimental techniques for constructing and exploring molecular
devices has inspired extensive theoretical study of charge transport in 
molecules, with potential application in molecular 
electronics.\cite{Nitzan2003,Reedbook,Tourbook,Cunibertibook}
One important issue related to stability of potential molecular devices
involves heating and heat transport in molecular junctions. 
This topic has attracted attention both experimentally\cite{1,2,3,4,5,6} and 
theoretically.\cite{7,8,9,10,11,12,13,14,15,16,heat}
Another closely related issue involves thermoelectric properties of such 
devices. While thermo-electricity in bulk is well studied,
corresponding measurements in molecular junctions were reported only 
recently.\cite{Poler,Majumdar}

Electron-vibration interactions in the junction (leading to inelastic effects
in charge transport\cite{review}) may cause junction heating and affect its 
heat transport properties.\cite{heat} 
Theoretical considerations of thermoelectric properties so far either 
completely disregarded such effects
(in treatments based on the Landauer theory\cite{Majumdar,Datta,Macia}) or
included it within a scattering theory framework.\cite{Segal,Walczak}
The latter treats the effect of vibrations as an inelastic electron 
scattering process, and changes in the non-equilibrium distributions
of electrons and vibrations are not described in a self-consistent manner.
It has been shown that such changes
may have qualitative effects on the transport.\cite{Dekker,strong}
Note also that scattering theory approaches may lead to erroneous
predictions due to neglecting effects of contacts' Fermi seas on the junction
electronic structure.\cite{Mitra}

This paper is motivated by recent measurements of the Seebeck
coefficient in molecular junctions.\cite{Majumdar} Considerations based on
Landauer formula were employed there to interpret experimental data.
Our goals here are: 1. to show the importance of vibrations for 
the Seebeck coefficient
and 2. to include vibrations in a fully self-consistent way within the
nonequilibrium Green's function approach for calculating the
Seebeck coefficient. The structure of the paper
is the following. Section~\ref{model} introduces the model, discusses 
methods used in calculations, and presents a simple (approximate) analytical
derivation to illustrate change in the Seebeck coefficient
expression (as compared to Landauer-based expression) 
when vibrations are taken into account. Section~\ref{results}
presents numerical results obtained in a fully self-consistent way.
Section~\ref{conclude} concludes. 

\section{\label{model}Model}
We consider a simple resonant-level model with the electronic level $|0>$
coupled to two electrodes left ($L$) and right ($R$) (each a free electron
reservoir at its own equilibrium).  The electron on the resonant level
(electronic energy $\varepsilon_0$) is linearly coupled to a single vibrational
mode (referred below as primary phonon) with frequency $\omega_0$. 
The latter is coupled to a phonon bath represented as a set of 
independent harmonic oscillators (secondary phonons). 
The system Hamiltonian is (here and below we use $\hbar=1$ and $e=1$)
\begin{align}
 \label{H}
 &\hat H = \varepsilon_0\hat c^\dagger\hat c +
 \sum_{k\in\{L,R\}} \varepsilon_k \hat c_k^\dagger\hat c_k +
 \sum_{k\in\{L,R\}} \left(V_k\hat c_k^\dagger\hat c + \mbox{h.c.}\right)
 \nonumber \\ &+
 \omega_0\hat a^\dagger\hat a +
 \sum_\beta\omega_\beta\hat b^\dagger_\beta\hat b_\beta +
 M_a\hat Q_a\hat c^\dagger\hat c +
 \sum_\beta U_\beta\hat Q_a\hat Q_\beta
\end{align}
where $\hat c^\dagger$ ($\hat c$) are creation (destruction) operators
for electrons on the bridge level, $\hat c_k^\dagger$ ($\hat c_k$) are
corresponding operators for electronic states in the contacts,
$\hat a^\dagger$ ($\hat a$)
are creation (destruction) operators for the primary phonon, and
$\hat b_\beta^\dagger$ ($\hat b_\beta$) are the corresponding operators for
phonon states in the thermal (phonon) bath.
$\hat Q_a$ and $\hat Q_\beta$ are phonon displacement operators
\begin{equation}
 \label{Q}
 \hat Q_a = \hat a + \hat a^\dagger \qquad
 \hat Q_\beta = \hat b_\beta + \hat b_\beta^\dagger
\end{equation}
The energy parameters $M_a$ and $U_\beta$ correspond to the vibronic
and the vibrational coupling respectively.
Eq.(\ref{H}) is often used as a generic model for describing effects
of vibrational motion on electronic conduction in molecular 
junctions.\cite{review}

After a small polaron (canonical or Lang-Firsov)
transformation\cite{Mahan,LangFirsov} the Hamiltonian (\ref{H})
takes the form (for details see Ref.\cite{strong})
\begin{eqnarray}
 \label{barH}
 \hat{\bar H} &=& \bar\varepsilon_0\hat c^\dagger\hat c +
 \sum_{k\in\{L,R\}} \varepsilon_k \hat c_k^\dagger\hat c_k +
 \sum_{k\in\{L,R\}} \left(V_k\hat c_k^\dagger\hat c\hat X_a+\mbox{h.c.}\right)
 \nonumber \\ &+&
 \omega_0\hat a^\dagger\hat a +
 \sum_\beta\omega_\beta\hat b^\dagger_\beta\hat b_\beta +
 \sum_\beta U_\beta\hat Q_a\hat Q_\beta
\end{eqnarray}
where
\begin{eqnarray}
 \label{bare0}
 \bar\varepsilon_0 &=& \varepsilon_0 - \Delta \qquad
 \Delta \approx \frac{M_a^2}{\omega_0}
 \\
 \label{Xa}
 \hat X_a &=& \exp\left[i\lambda_a\hat P_a\right] \qquad
 \lambda_a = \frac{M_a}{\omega_0}
\end{eqnarray}
$\Delta$ is the electron level shift due to coupling to the primary phonon
and $\hat X_a$ is primary phonon shift generator. 
$\hat P_a=-i(\hat a-\hat a^\dagger)$ is the phonon momentum operator.

The mathematical quantity of interest is the single electron Green function 
(GF) on the Keldysh contour
\begin{equation}
 \label{defGFKeldysh}
 G(\tau_1,\tau_2) \equiv -i<T_c\,\hat c(\tau_1)\,\hat c^\dagger(\tau_2)>_H
\end{equation}
Following Ref.~\cite{strong}, we approximate it by
\begin{align}
 \label{appGFKeldysh}
 G(\tau_1,\tau_2) &\approx  
 -\frac{i}{\hbar}<T_c \hat c(\tau_1)\,\hat c^\dagger(\tau_2)>_{\bar H}\,
 \times <\hat X_a(\tau_1)\,\hat X_a^\dagger(\tau_2)>_{\bar H}
 \nonumber \\
 &\equiv
 G_c(\tau_1,\tau_2)\,{\cal K}(\tau_1,\tau_2)
\end{align}
where $G_c(\tau_1,\tau_2)$ is the pure electronic GF 
while ${\cal K}(\tau_1,\tau_2)$ corresponds to the Franck-Condon factor. 
In Ref.~\cite{strong} we have developed a self-consistent scheme for 
evaluating this functions, leading to the coupled set of equations 
\begin{align}
\label{XXKeldysh}
 &{\cal K}(\tau_1,\tau_2) =
 \exp\left\{\lambda_a^2\left[i\hbar D_{P_aP_a}(\tau_1,\tau_2)
                -<\hat P_a^2>\right]\right\}
 \\
  \label{Dyson}
 &D_{P_aP_a}(\tau,\tau') = D_{P_aP_a}^{(0)}(\tau,\tau')
 \\
 &+ \int_c d\tau_1 \int_c d\tau_2\, D_{P_aP_a}^{(0)}(\tau,\tau_1)\,
 \Pi_{P_aP_a}(\tau_1,\tau_2)\, D_{P_aP_a}(\tau_2,\tau')
 \nonumber \\
 \label{GcKeldysh}
 &G_c(\tau,\tau') = G_c^{(0)}(\tau,\tau')
 \\
 &+ \sum_{K=\{L,R\}}\int_c d\tau_1 \int_c d\tau_2\, G_c^{(0)}(\tau,\tau_1)\,
 \Sigma_{c,K}(\tau_1,\tau_2)\, G_c(\tau_2,\tau')
 \nonumber
\end{align}
where $D_{P_aP_a}(\tau,\tau')\equiv -i<T_c \hat P_a(\tau)\,\hat P_a(\tau')>$ 
is the phonon GF, $D_{P_aP_a}^{(0)}$ and $G_c^{(0)}$ are the phonon and 
electron Green functions when the two sub-systems are uncoupled ($M_a=0$).
The functions $\Pi_{P_aP_a}$ and $\Sigma_{c,K}$ in 
Eqs.~(\ref{Dyson}) and (\ref{GcKeldysh}) are given by
\begin{align}
 \label{DSEKeldysh}
 &\Pi_{P_aP_a}(\tau_1,\tau_2) = \sum_\beta |U_\beta|^2
 D_{P_\beta P_\beta}(\tau_1,\tau_2)
 - i\lambda_a^2\sum_{k\in\{L,R\}}|V_k|^2
 \nonumber \\ &\times
 \left[
     \hbar g_k(\tau_2,\tau_1)\,G_c(\tau_1,\tau_2)\,{\cal K}(\tau_1,\tau_2)
     +  (\tau_1\leftrightarrow\tau_2)\right]
  \\
 \label{GcSEKeldysh}
 &\Sigma_{c,K}(\tau_1,\tau_2) = \sum_{k\in K} |V_k|^2 g_k(\tau_1,\tau_2)
 <T_c \hat X_a(\tau_2)\hat X_a^\dagger(\tau_1)>
\end{align}
These functions play here a role similar to self-energies in standard
many-particle theory.
Here $K=L,R$ and $g_k$ is the free electron Green function for state $k$
in the contacts. For details of derivation see Ref.\cite{strong}.
A self-consistent solution scheme implies solving 
Eqs.~(\ref{XXKeldysh})-(\ref{GcSEKeldysh}) iteratively until convergence.
As a convergence parameter we used population of the level 
$n_0=<\hat c_0^\dagger\hat c_0>$. When $n_0$ for subsequent steps of the
iterative cycle differed by less than a predefined tolerance 
(taken in the calculations below to be $10^{-6}$), 
convergence was assumed to be achieved.

Once the electron GF (\ref{appGFKeldysh}) is obtained, its lesser and greater
projections are used to get steady-state current through the 
junction\cite{current,HaugJauho} 
\begin{equation}
\label{IK}
 I_K = \left<\hat I_K\right> = \int\frac{dE}{2\pi}\,
 \left[\Sigma_K^{<}(E)\, G^{>}(E) - \Sigma_K^{>}(E)\, G^{<}(E)\right]
\end{equation}
at interface $K=L,R$. Here
\begin{eqnarray}
 \label{SEKlt}
 \Sigma_K^{<}(E) &=& i f_K(E) \Gamma_K(E) \\
 \label{SEKgt}
 \Sigma_K^{>}(E) &=& -i [1-f_K(E)] \Gamma_K(E)
\end{eqnarray}
with $f_K(E)=[\exp(\beta(E-\mu_K))+1]^{-1}$ the Fermi distribution 
in the contact $K=L,R$ and
\begin{equation}
 \label{GammaK}
 \Gamma_K(E) = 2\pi \sum_{k\in K} |V_k|^2 \delta(E-\varepsilon_k)
\end{equation}

The Seebeck coefficient is defined by
\begin{equation}
 \label{defSeebeck}
 S(I) = \frac{V(I)}{\Delta T(I)}
\end{equation}
where $V(I)$ is the voltage bias that yields current $I$ at $\Delta T=0$,
while $\Delta T(I)$ is the temperature difference between contacts that 
yields the same current at $V=0$. The linear regime corresponds to the 
$I\to 0$ limit of Eq.(\ref{defSeebeck}).

Below we present calculations of the Seebeck coefficient using different
levels of approximations. In particular we compare results of a
simple scattering theory-like approach and a
full self-consistent calculation based on the procedure described above.
The simple approach is essentially a first step of the full self-consistent 
iterative solution with the additional assumption of no coupling
to the thermal bath for the molecular vibration 
($U_\beta \to 0$ in Eq.(\ref{barH})).

\section{\label{transport_coef}Transport coefficients}
Before presenting results of numerical calculations, we point out how
transport coefficients are introduced.  In the 
Landauer regime of transport (electron-phonon interaction
disregarded, both carriers scatter ballistically),
the electric and thermal fluxes, $I$ and $J$, 
are given by\cite{Datta,Butcher,heat}
\begin{equation}
 \label{ILandauer}
 \begin{aligned}
  I &= \frac{2|e|}{\hbar}\int_{-\infty}^{+\infty}\frac{dE}{2\pi}\,
       {\cal T}_{el,0}(E)[f_L(E)-f_R(E)] 
  \\
  J &= \frac{2}{\hbar}\int_{-\infty}^{+\infty}\frac{dE}{2\pi}\,
       (E-E_F){\cal T}_{el,0}(E)[f_L(E)-f_R(E)] 
  \\
    &+ \frac{1}{\hbar}\int_{0}^{\infty}\frac{d\omega}{2\pi}\,
       \omega {\cal T}_{ph,0}(\omega)[N_L(\omega)-N_R(\omega)]
 \end{aligned}
\end{equation}
where $E_F$ is a common Fermi energy in the absence of bias,
\begin{equation}
 \begin{aligned}
  {\cal T}_{el,0}(E) &= \mbox{Tr}[\Gamma_L(E)\,G^r(E)\,\Gamma_R(E)\,G^a(E)]
  \\
  {\cal T}_{ph,0}(\omega) &= \mbox{Tr}[\Omega_L(E)\,D^r(E)\,\Omega_R(E)\,D^a(E)]
 \end{aligned}
\end{equation}
are the electron and phonon transmission coefficients 
in the absence of electron-phonon coupling, and where
\begin{equation}
 \label{Omega}
 \Omega_K^{ph}(\omega) \equiv 2\pi\sum_{\beta\in K} 
 |U_\beta|^2\delta(\omega-\omega_\beta)
 \qquad K=L,R
\end{equation}
is the broadening of the molecular vibration due to coupling to its 
thermal environment. In the linear response regime the currents are 
linear in the applied driving forces -- the bias $V$ and the temperature 
difference $\Delta T$ 
\begin{equation}
 \begin{aligned}
  I &= G\cdot V + L\cdot \Delta T \\
  J &= R\cdot V + F\cdot \Delta T
 \end{aligned}
\end{equation}
Here $G$ and $F$ are the electrical and thermal conductions, respectively,
and $L$ is known as the thermoelectric coefficient.
The coefficients are given by\cite{Datta,Butcher}
\begin{align}
 \label{G}
 G &= -\frac{e^2}{\pi\hbar}\int_{-\infty}^{+\infty} dE\,
      [-\beta f'(E)]{\cal T}_{el,0}(E) 
 \\
 \label{L}
 L &= -\frac{|e|}{\pi\hbar}\int_{-\infty}^{+\infty} dE\,
      [-\beta f'(E)]{\cal T}_{el,0}(E)\frac{E-E_F}{T}
 \\
 \label{R}
 R &= -\frac{|e|}{\pi\hbar}\int_{-\infty}^{+\infty} dE\,
      [-\beta f'(E)]{\cal T}_{el,0}(E) (E-E_F) = L\cdot T
 \\
 \label{S}
 F &= \frac{1}{\pi\hbar}\int_{-\infty}^{+\infty} dE\,
      [-\beta f'(E)]{\cal T}_{el,0}(E) \frac{(E-E_F)^2}{T}
 \nonumber \\
   &+ \frac{1}{2\pi\hbar}\int_0^\infty d\omega[-\beta N'(E)]
      {\cal T}_{ph,0}(\omega)\frac{\omega^2}{T}
\end{align}
where $f'(E)$ is derivative of Fermi-Dirac distribution, 
$N'(\omega)$ is derivative of Bose-Einstein distribution,
$T$ is temperature ($\beta=1/T$), and $E_F$ is the Fermi energy in the leads.
Note the existence of Onsager relation, $L\cdot T = R$, between the 
cross coefficients. Note also that the coefficient $F$ in (\ref{S}) 
contains two contributions, one corresponding to energy transfer by electrons,
the other -- by phonons. For discussion of the additive from of $F$ and 
the relative importance of these contributions see 
Ref.~\cite{heat}. The Seebeck coefficient is given in terms of these transport 
coefficients by
\begin{equation}
 \label{Seebeck_linear}
 S = \frac{L}{G}
\end{equation}
Below we focus on these two coefficients -- $G$ and $L$ -- only.
Making the approximation $[-\beta f'(E)]\approx\delta(E-E_F)$ in (\ref{G}),
and utilizing the Sommerfield expansion\cite{Landau} 
in (\ref{L}), Eq.(\ref{Seebeck_linear}) leads to 
\begin{equation}
 S = \frac{\pi^2k_b^2T}{3|e|}\frac{\partial\ln {\cal T}_{el,0}(E)}{\partial E} 
\end{equation}
which is Eq.(4) of Ref.~\cite{Datta}.

\section{\label{results}Calculation of the Seebeck coefficient}
As discussed in Section~\ref{model}, the simplest calculation that takes
into account the electron-vibration interaction term (the $M_a$ term
in Eq.(\ref{H})) corresponds to inelastic scattering of the transmitted
electron from the phonon at the given initial temperature. Within the
self-consistent scheme presented in Section~\ref{model} this result is
obtained after the first iteration step, with the influence of
the (free) vibration on the electronic GF is taken into account but not 
vice versa. For this reason the coupling $U_\beta$ to the thermal bath 
can be disregarded. For the model (\ref{barH}) 
this calculation yields\cite{footnote} 
\begin{align}
 \label{Isimple}
 I &= \frac{|e|}{\pi\hbar}\sum_{n,m=-\infty}^{+\infty}
      I_n(2\lambda^2\sqrt{N_0(N_0+1)})I_m(2\lambda^2\sqrt{N_0(N_0+1)})
 \nonumber \\ &\times 
      e^{\beta(n+m)\omega_0/2-2\lambda^2(2N_0+1)}
      \int_{-\infty}^{+\infty}dE\, {\cal T}_{el,0}(E)
 \\ &\times
      \left\{f_L(E+n\omega_0)[1-f_R(E-m\omega_0)]
           -f_R(E+m\omega_0)[1-f_L(E-n\omega_0)]\right\}
 \nonumber
\end{align}
where $N_0=[e^{\beta\omega_0}-1]^{-1}$ and $I_n$ is the modified 
Bessel function of order $n$.\cite{AS} 
For $M_a=0$ Eq.(\ref{Isimple}) reduces back to (\ref{ILandauer}).
Linearization in the bias potential $V=V_L-V_R$ and in the temperature
difference $\Delta T=T_L-T_R$ leads to the phonon-renormalized 
transport coefficients
\begin{align}
 \label{Gph}
 G &= -\frac{e^2}{\pi\hbar} \sum_{n,m=-\infty}^{+\infty}
      I_n(2\lambda^2\sqrt{N_0(N_0+1)})I_m(2\lambda^2\sqrt{N_0(N_0+1)})
 \nonumber \\ &\times 
      e^{\beta(n+m)\omega_0/2-2\lambda^2(2N_0+1)}
      \int_{-\infty}^{+\infty}dE\, {\cal T}_{el,0}(E)
 \\ &\times
      \left\{[-\beta f'(E-m\omega_0)]f(E+n\omega_0)
      +[-\beta f'(E+m\omega_0)][1-f(E-n\omega_0)]\right\}
 \nonumber \\
 \label{Lph}
 L &= -\frac{|e|}{\pi\hbar} \sum_{n,m=-\infty}^{+\infty}
      I_n(2\lambda^2\sqrt{N_0(N_0+1)})I_m(2\lambda^2\sqrt{N_0(N_0+1)})
 \nonumber \\ &\times
      e^{\beta(n+m)\omega_0/2-2\lambda^2(2N_0+1)}
      \int_{-\infty}^{+\infty}dE\, {\cal T}_{el,0}(E)
 \nonumber \\ &\times
      \left\{[-\beta f'(E-m\omega_0)]f(E+n\omega_0)\frac{E-m\omega_0-E_F}{T}
 \right. \\ &\quad \left.
      +[-\beta f'(E+m\omega_0)][1-f(E-n\omega_0)]\frac{E+m\omega_0-E_F}{T}
      \right\}
 \nonumber
\end{align}

Using Eqs.~(\ref{Gph}) and (\ref{Lph}), the Seebeck coefficient
is calculated from Eq.(\ref{Seebeck_linear}). 
To this end we first calculate the currents $I(V,\Delta T=0)$
and $I(V=0,\Delta T)$ as functions of $V$ and $\Delta T$.
The inverted functions $V(I,\Delta T=0)$ and $\Delta T(I,V=0)$
are then used in (\ref{Seebeck_linear}) to yield $S(I)$
(expressed below as $S(V)$ with $V=V(I,\Delta T=0)$).
In the calculations presented below we have used symmetric bias and 
temperature differences across the junction: $\mu_{L,R}=E_F\pm V/2$,
$T_{L,R}=T\pm\Delta T/2$.

\begin{figure}[htbp]
\centering\includegraphics[width=0.6\linewidth]{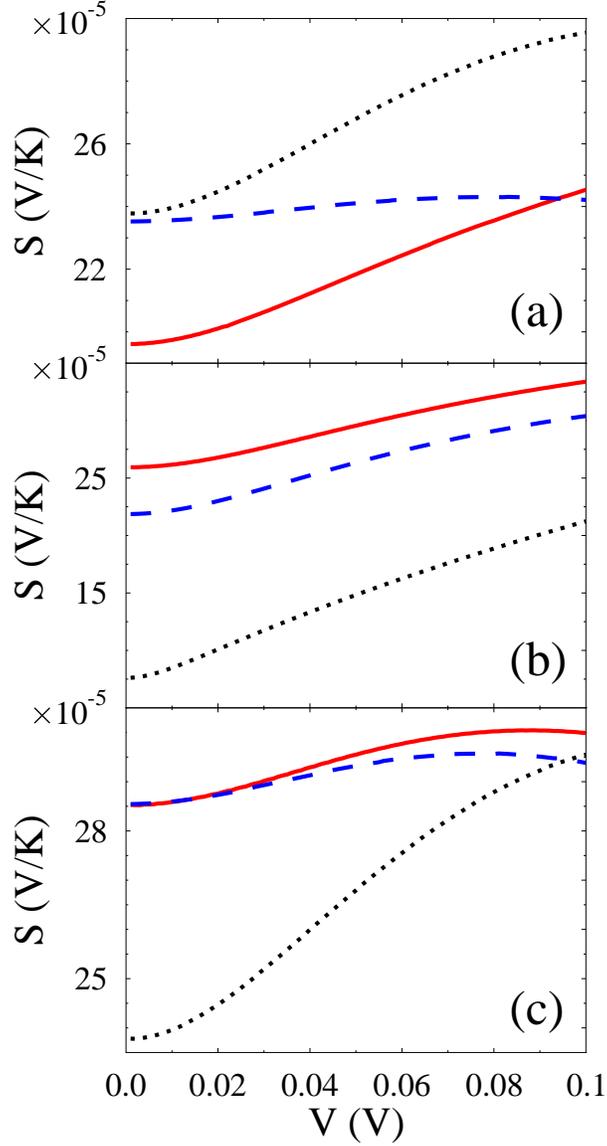}
\caption{\label{fig_get_seebeck_iv}(Color online)
Seebeck coefficient vs.  applied bias. 
Shown are results of full self-consistent 
calculation (solid line, red), 
scattering theory approach based on model (\ref{Isimple}) 
(dashed line, blue), and elastic scattering case (dotted line, black)
calculated with (a) `standard' set of parameters (see text for parameters),
(b) higher $\varepsilon_0-E_F$ gap, and (c) weaker electron-phonon
coupling $M_a$.
}
\end{figure}

Figure~\ref{fig_get_seebeck_iv}a shows the Seebeck coefficient $S$
as a function of the bias potential, calculated at
$T=300$~K using the energetic parameters $E_F=0$, $\varepsilon_0=0.2$~eV,
$\Gamma_L=\Gamma_R=0.005$~eV, $\omega_0=0.05$~eV, $M_a=0.1$~eV, 
and $\Omega^{ph}_L=\Omega^{ph}_R=0.005$~eV. 
The last is a wide-band approximation 
for molecular vibration broadening (\ref{Omega}) due to coupling to thermal 
baths (for detailed discussion see Ref.~\cite{weak}).
Figures~\ref{fig_get_seebeck_iv}b and c show similar results for
a higher $\varepsilon_0-E_F$ gap and a weaker electron-phonon coupling
$M_a$, respectively.
To show the effect of electron-phonon coupling on the Seebeck coefficient
we compare the results obtained from the full self-consistent calculation
and the scattering theory approximation to the elastic $M_a=0$ limit.

\begin{figure}[htbp]
\centering\includegraphics[width=\linewidth]{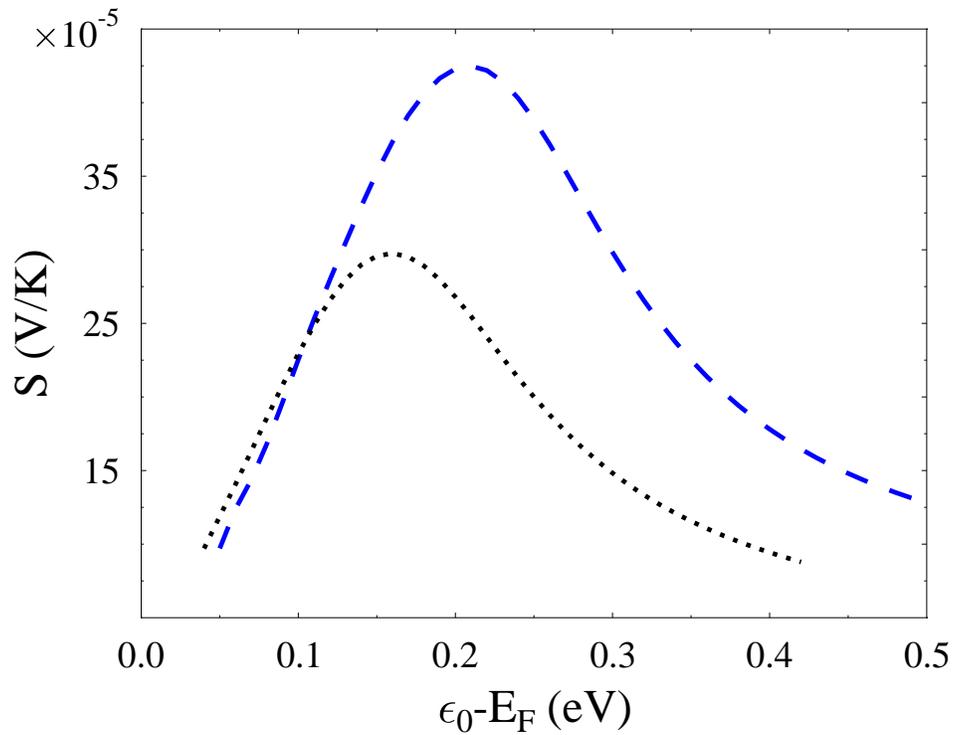}
\caption{\label{fig_get_seebeck_ie}(Color online)
Seebeck coefficient vs. energy position of the molecular level
for the model of Eq.(\ref{Isimple}).
Shown are results with (dashed line, blue) and without (dotted line, black)
electron-phonon interaction. 
Calculation is done at $V=0.05$~V. Other parameters are the same as in 
Fig.~\ref{fig_get_seebeck_iv}
}
\end{figure}

\begin{figure}[htbp]
\centering\includegraphics[width=\linewidth]{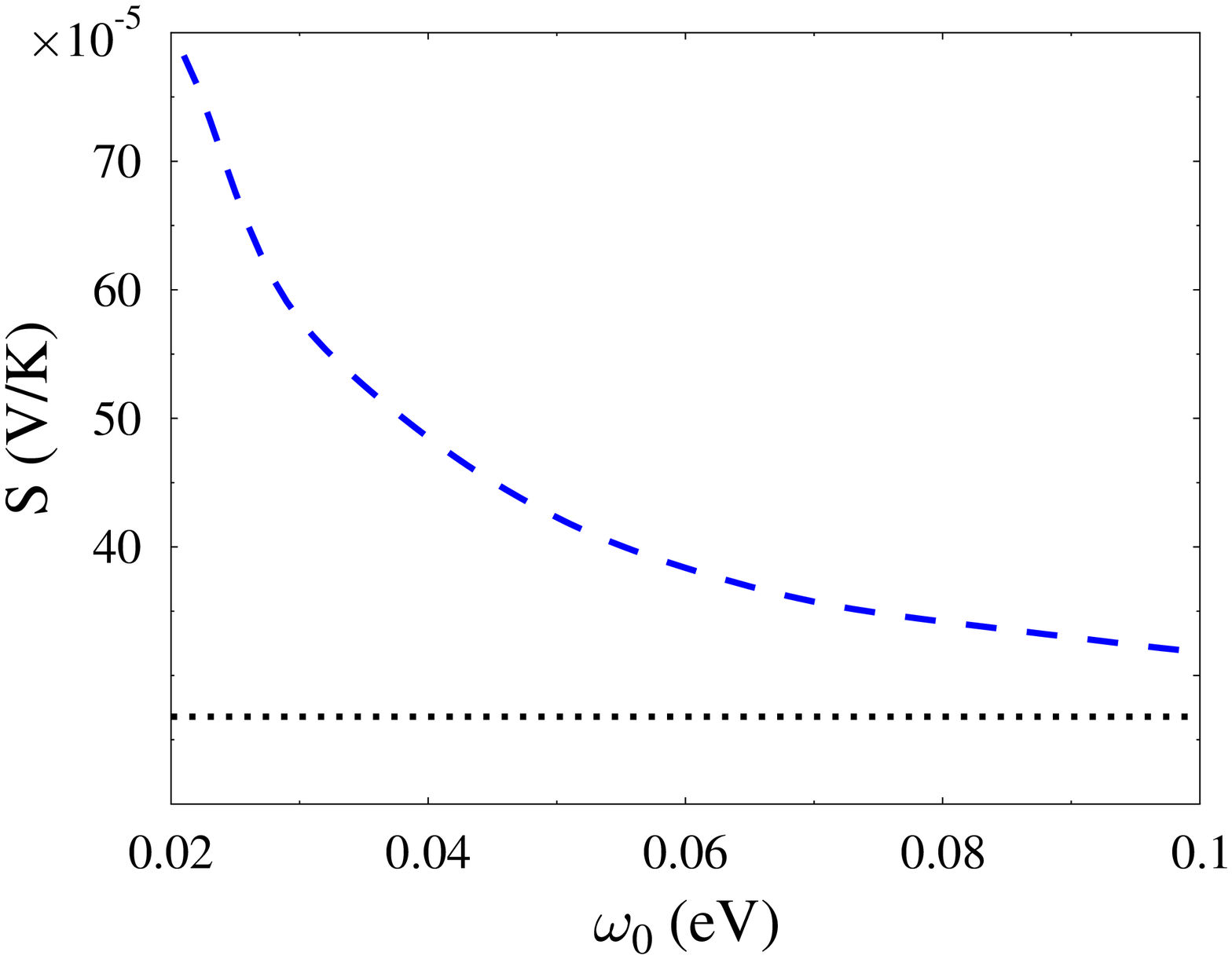}
\caption{\label{fig_get_seebeck_iw}(Color online)
Seebeck coefficient vs. vibrational mode frequency
for the model of Eq.(\ref{Isimple}).
Shown are result with (dashed line, blue) and without (dotted line, black)
electron-phonon interaction.
Calculation is done at $V=0.05$~V. Other parameters are the same as in
Fig.~\ref{fig_get_seebeck_iv}
}
\end{figure}

\begin{figure}[htbp]
\centering\includegraphics[width=\linewidth]{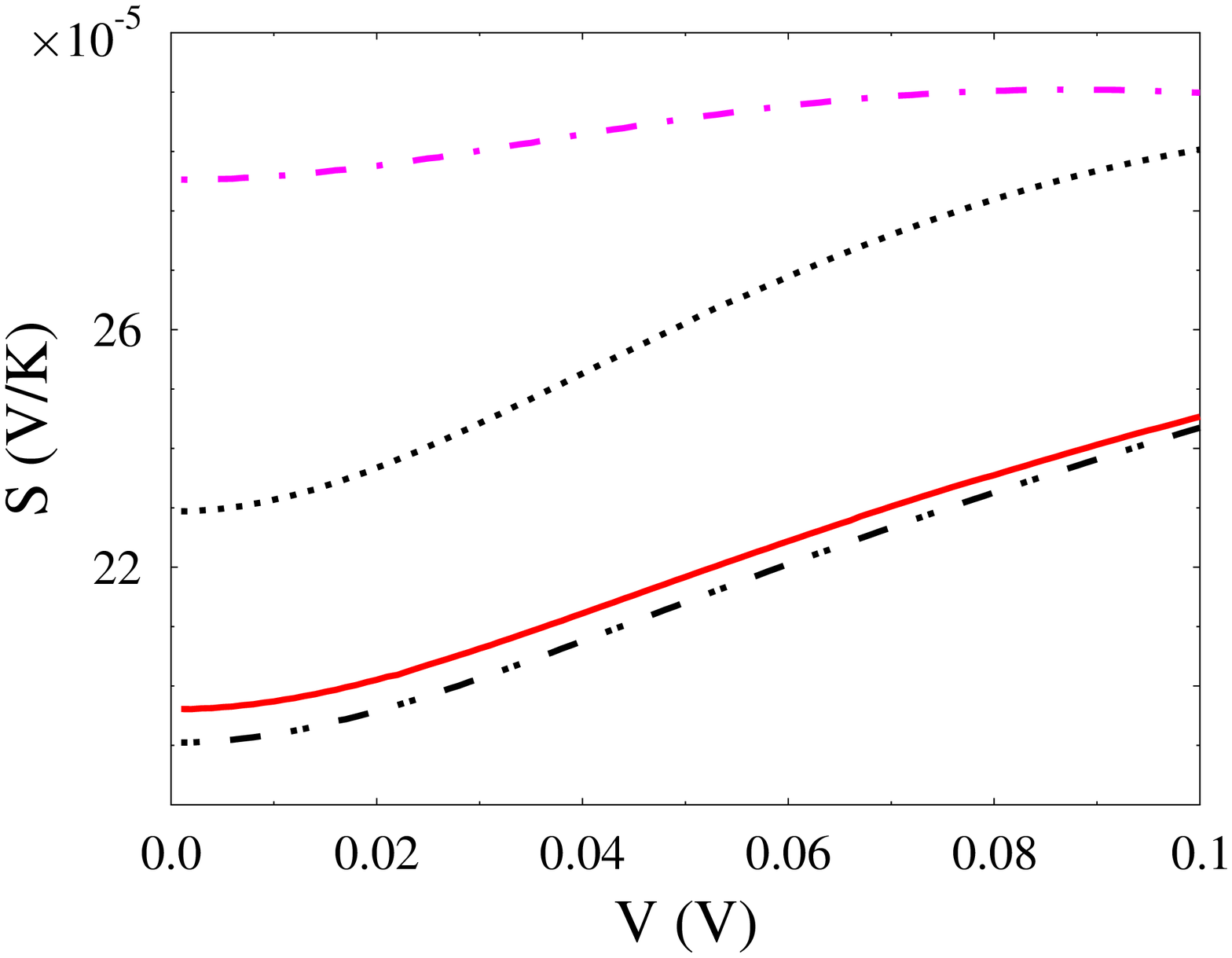}
\caption{\label{fig_s_ex_w05_gphxxx}(Color online)
Seebeck coefficient vs. applied bias for model described in Section~\ref{model}.
Solid line (red) and dash-dotted line (magenta) are identical 
to solid lines (red) 
in Figs.~\ref{fig_get_seebeck_iv}a and c respectively. 
Also shown result for $\Omega^{ph}_L=\Omega^{ph}_R=0.002$~eV 
(dash-double dotted line, black).
Elastic case (dotted line, black) is shown for comparison.
}
\end{figure}

In the following figures we consider the inelastic effects only within
the scattering theory approximation (which requires a far smaller numerical
effort). The dependence of $S$ on the energy gap  $\varepsilon_0-E_F$
is shown in Fig.~\ref{fig_get_seebeck_ie}, and its variation as function
of $\omega_0$ is displayed in Fig.~\ref{fig_get_seebeck_iw}.
In these figures $V$ is kept at the value $0.05$~V and all unvaried parameters
are the same as in Fig.~\ref{fig_get_seebeck_iv}. 
Finally, in Fig.~\ref{fig_s_ex_w05_gphxxx} we compare self-consistent
results (some of them already shown in Fig.~\ref{fig_get_seebeck_iv})
obtained for different choices of electron-phonon coupling $M_a$ and
vibrational broadening $\Omega^{ph}$.

The following observations can be made regarding these results:
\begin{enumerate}
\item In contrast to inelastic tunneling features usually seen
in the second derivative $d^2I/dV^2$ of the current-voltage
characteristic near $|eV|=\hbar\omega_0$ no such threshold behavior
is seen in Fig.~\ref{fig_get_seebeck_iv}.
This lack of threshold behavior in the inelastic contribution
to the Seebeck coefficient results the fact that thermoelectric 
conduction is assosiated with the tails of the lead Fermi-Dirac distributions,
and these tails wash away any thershold structure.
\item Inelastic contributions can have a substantial effect on the
Seebeck coefficient and its voltage dependence, however the assessment
of these contributions is sensitive to the approximation used and cannot
generally be based on the scattering theory level of calculation.
Indeed, as seen in Figure~\ref{fig_get_seebeck_iv} $S(V)$ behavior 
may change qualitatively upon going from the scattering to the self-consistent
calculation.
\item Focusing on the self-consistent results, Figs.~\ref{fig_get_seebeck_iv}
and \ref{fig_s_ex_w05_gphxxx} show that the inelastic effect on the Seebeck
coefficient can be positive or negative, depending on the other energetic
parameters in the system. A change of sign as a function of $\varepsilon_0$
is seen also in the scattering theory results of Fig.~\ref{fig_get_seebeck_ie}.
The existence of a similar crossover in the self-consistent results
can be inferred by comparing Figures~\ref{fig_get_seebeck_iv}a and b. 
\item A smaller value of the electron-phonon coupling $M_a$ naturally
leads to a smaller difference between self-consistent and scattering theory
calculations (compare Figs.~\ref{fig_get_seebeck_iv}a and c). Still the 
coupling chosen in Fig.~\ref{fig_get_seebeck_iv}c is strong enough 
to give appreciable difference between the inelastic and elastic results. 
\item In Fig.~\ref{fig_get_seebeck_ie} the Seebeck coefficient is seen
to go through a maximum as a function of the gap $\varepsilon_0-E_F$,
with inelastic contribution affecting the position and height of 
the observed peak. The behavior seen in Fig.~\ref{fig_get_seebeck_ie}
can be rationalized by noting that for, say, 
$T_L>T_R$ electrons with energies $E>E_F$ contribute most
to left-to-right current, while those with $E<E_F$ dominate
right-to-left current.
This gives no thermoelectric current when $\varepsilon_0=E_F$,
hence as $\varepsilon_0\to E_F$ one needs a higher temperature difference
in order to compensate for the same bias. As a result, the Seebeck coefficient
goes down at $\varepsilon_0-E_F\to 0$. On the other hand when
$\varepsilon_0-E_F\gg \Gamma, k_BT$ the two contributions cancel each other,
and hence Seebeck coefficient drops down once more.
\item The dependence of $S$ on $\omega_0$ (Fig.~\ref{fig_get_seebeck_iw}
demonstrates it within a scattering theory level calculation) is in line
with the expectation that $S$ should attain its classical limit as the 
vibration becomes more rigid.
\item We have found (not shown) that effects on $S$ of varying the 
electronic ($\Gamma$) or vibrational ($\Omega$) widths show a similar 
trend as varying the gap $\varepsilon_0-E_F$ or the vibrational frequency
$\omega_0$, respectively.
\end{enumerate}

\section{\label{conclude}Conclusion}
We studied the influence of molecular vibration (inelastic effects) 
on the Seebeck coefficient for molecular junction transport using a 
simple model of one molecular level (representing the participating 
molecular state) coupled to two contacts and to one molecular vibration. 
Two approaches to the model were considered:
a simplified scattering model represented by Eq.(\ref{Isimple}) and the full 
self-consistent approach described in Section~\ref{model}.
The simplified approach ignores the mutual influence
of electronic and vibrational subsystems. Note, the structure of 
the expression for current, Eq.(\ref{Isimple}) is just a difference between
two scattering fluxes (left-to-right minus right-to-left).
Results of simplified model calculation are compared to
the full self-consistent approach where both mutual electron-vibration
influence and vibration coupling to thermal baths are taken into account
(for detailed description of the approach see Ref.~\cite{strong}).
We show that inelastic effects have non-negligible influence on the
resulting Seebeck coefficient for the junction, for a reasonable
choice of parameters. Electron-vibration interaction can either
increase or decrease Seebeck coefficient depending on the physical situation.
We study the dependence of this influence on different parameters
of the model (applied bias, gap between molecular level and Fermi energy, 
strengths of coupling between molecule and contacts, between tunneling
electron and molecular vibration, between
molecular vibration coupling and thermal baths, and frequency of the vibration).
Comparing results of the two approaches, we show that scattering theory
based approach may fail both quantitatively and qualitatively.
Experimental data presented in Ref.~\cite{Majumdar} does not give
conclusive evidence for the relative importance of inelastic processes.
More extensive measurements showing dependence of the Seebeck coefficient
on junction parameters are needed in order to make definite conclusion.
In particular, isotopic effects should influence vibration-related
part of the Seebeck coefficient (see Figs.~\ref{fig_get_seebeck_iw}
and \ref{fig_s_ex_w05_gphxxx}).
Change in electron-phonon coupling would also reveal the inelastic part of the 
Seebeck coefficient (see \ref{fig_s_ex_w05_gphxxx}). 
Finally, results of our model calculations 
done with a reasonable set of parameters provide the Seebeck coefficient
to be of the order of $\sim 10^{-4}$~V/K. Results reported in 
Ref.~\cite{Majumdar} yield $S\sim 10^{-5}$~V/K. Since molecules
used in the experiment\cite{Majumdar} are characterized by relatively
big gap $\varepsilon_0-E_F$, our results are in good agreement with
the measured data. Indeed, for $\varepsilon_0-E_F\sim 1$~eV calculated
Seebeck coefficient (see Fig.~\ref{fig_get_seebeck_ie}) 
becomes of the experimentally observed order of magnitude.

\section*{Acknowledgments}
M.G. and M.A.R. are grateful to the DARPA MoleApps program, to NSF-MRSEC,
NSF-Chemistry, and NASA-URETI programs for support.
The research of A.N. is supported by the Israel Science Foundation,
the US-Israel Binational Science Foundation, and the Germany-Israel 
Foundation. This paper is dedicated to Raphy Levine -- a friend, colleague
and a pioneer of our field.


\begin{thebibliography}{99}
\bibitem{footnote}To obtain Eq.(\ref{Isimple}) from Eq.(\ref{IK}) 
the "first iterative step" of the self consistent procedure\cite{strong} 
has to be taken. This leads to (see Eq.(\ref{appGFKeldysh}))
\[
 G^{>,<}(t) = G_c^{>,<}(t)\,{\cal K}^{>,<}(t)
\] 
In order to keep the the $L\leftrightarrow R$ symmetry for the current 
expression, one has to introduce the second FC factor in the Keldysh equation
for $G_c$
\[
 G_c^{>,<}(t) = \sum_{K=L,R}
 \int_{-\infty}^{+\infty}dt_1 \int_{-\infty}^{+\infty}dt_2\,
 G_c^r(t-t_1)\,\Sigma_K^{>,<}(t_1-t_2)\,{\cal K}^{<,>}(t_2-t_1)\,G^a(t_2)
\]
Substituting these expressions into Eq.(\ref{IK}) and expanding FC factors in
term of Bessel functions leads to Eq.(\ref{Isimple}).
\end{thebibliography}
\end{document}